\documentclass[preprint,aps,groupedaddress,tightenlines]{revtex4}
\usepackage[american]{babel}
\usepackage{color,graphicx,amsmath,amssymb,bm,epsfig}
\begin{document}
\newcommand\ket[1]{\left|#1\right\rangle}
\newcommand\bra[1]{\left\langle#1\right|}
\newcommand\lavg{\left\langle}
\newcommand\ravg{\right\rangle}
\renewcommand{\theequation}{\arabic{equation}}
\def\bra#1{\left\langle{#1}\,\right|\,}    
\def\ket#1{\,\left|\,{#1}\right\rangle}    

\title{Clauser-Horne inequality for electron counting
        statistics in multiterminal mesoscopic conductors}

\author{Lara Faoro$^{(1)}$, Fabio Taddei$^{(1,3)}$ and Rosario Fazio$^{(3)}$}

\affiliation{$^{(1)}$ISI Foundation, Viale Settimio Severo,
                65, I-10133  Torino, Italy}
\affiliation{$^{(2)}$Department of Physics and Astronomy, Rutgers University,
		136 Frelinghuysen Road, Piscataway, New Jersey 08854 USA}
\affiliation{$^{(3)}$NEST-INFM $\& $ Scuola Normale
                Superiore, I-56126 Pisa, Italy}

\date{\today}

\begin{abstract}
In this paper we derive the Clauser-Horne (CH) inequality for the full 
electron counting statistics in a mesoscopic multiterminal conductor
and we discuss its properties.
We first consider the idealized situation in which a flux of
entangled electrons is generated by an 
{\em entangler}.
Given a certain average number of incoming entangled electrons,
the CH inequality can be evaluated for different numbers of
transmitted particles.
Strong violations occur when the number of transmitted charges on the two
terminals is the same ($Q_1=Q_2$), whereas no violation is found for $Q_1\ne Q_2$.
We then consider two actual setups that can be realized experimentally.
The first one consists of a three terminal normal beam splitter and the second one of
a hybrid superconducting structure.
Interestingly, we find that the CH inequality is violated for the
three terminal normal device. 
The maximum violation scales as $1/M$ and $1/M^2$ for the entangler and
normal beam splitter, respectively, $2M$ being the average number of
injected electrons.
As expected, we find full violation of the CH inequality in the case of the
superconducting system.
\end{abstract}

\maketitle

\section{Introduction}

Entanglement~\cite{bell87} denotes the nonlocal correlations that
exist, even in the absence of direct interaction, between two (spatially 
separated) parts of a given quantum system. Since the early days of quantum 
mechanics, understanding the phenomenon of entanglement has been central to the 
understanding of the foundations of quantum theory.  Besides its fundamental 
importance, a great deal of interest has been brought forth by its role in quantum 
information~\cite{nielsen00}. Entanglement is believed to be the main ingredient 
of computational speed-up in quantum information protocols. 

Most of the work on entanglement has been performed in optical systems 
with photons~\cite{zeilinger99}, cavity QED systems~\cite{rauschenbeutel00} and 
ion traps~\cite{sackett00}. Only recently attention has been devoted
to the manipulation of entangled states in a solid state environment.
This interest, originally motivated by the idea to realize a solid state 
quantum computer~\cite{makhlin01,awschalom02}, has been rapidly growing and by 
now several works discuss how to generate, manipulate and detect entangled states
in solid state systems. It is probably worth to emphasize already 
at this point that, differently from the situation encountered in quantum optics, in 
solid state system entanglement is rather common. What is not trivial is 
its control and detection (especially if the interaction 
between the different subsystems forming the entangled state is
switched off).

Despite the large body of knowledge developed in the study of optical systems,
new strategies have to 
be designed to reveal the signatures of non-local correlations in the case of 
electronic states. For mesoscopic conductors, the prototype scheme was discussed in 
Ref.~\cite{burkard00}. In this work it has been shown  that the presence of spatially 
separated pairs of entangled electrons, created by some {\em entangler}, can be revealed 
by using a beam splitter and by measuring the correlations of the current fluctuations
in the leads. Provided that the  electrons injected are in an entangled state 
bunching and anti-bunching behavior for the cross-correlations of current fluctuations
are found depending on whether the state is a spin singlet or a spin triplet.
Not only the noise, 
but the full counting statistics is sensitive to the presence of entanglement in the 
incoming beam~\cite{taddei02}. The distribution of transmitted electrons is binomial and  
symmetric with respect to the average number of transmitted charges. Moreover, this is
important for the problem studied in the present work,  the joint probability for 
counting electrons at different leads unambiguously characterizes 
the state of the incident electrons if one uses  spin-sensitive electron 
counters.
In this case the joint probability cannot be 
expressed as a product of single-terminal probabilities.

Given the general setup to detect entanglement an important issue is 
to understand how to generate it. This has been discussed in several 
papers. Most  of the existing proposals are based on the generation of
Bell states by means of electron-electron interaction. This can be 
achieved through superconducting correlations~\cite{footnote} in hybrid normal - 
superconducting~\cite{loss00,loss01,lesovik01,samuelsson03} and superconductor - 
carbon nanotubes systems~\cite{bena01,bouchiat02}, 
quantum dots in the Coulomb blockade regime~\cite{saraga03}
or Kondo-like impurities~\cite{costa01}. Then, by 
using energy or spin filters, the two electrons forming the Bell state are separated.
The entanglement can be created in the spin or in the 
orbital \cite{samuelsson03} degrees of freedom. Very recently, as it is also discussed 
in Section~\ref{ns}, 
it was shown that in a mesoscopic multi-terminal conductor entanglement can be produced 
also in absence of electron interaction~\cite{beenakker03}. 
Besides electrons, it is possible to produce entangled states with Cooper pairs in 
superconducting nanocircuits~\cite{plastina01} or by coupling a mesoscopic Josephson 
junctions with superconducting resonators~\cite{buisson00,marquardt01,you02,plastina03}.

Since Bell's work~\cite{bell64}, it is known that a classical theory formulated in 
terms of a hidden variables satisfying reasonable condition of locality, yields 
predictions which are different from those of quantum mechanics. These predictions were 
casted into the form of inequalities which any realistic local theory must obey. 
Bell inequalities have been formulated for mesoscopic multi-terminal conductors 
in Refs.~\cite{kawabata00,chtchelkatchev02,samuelsson03} in terms of electrical noise 
correlations at different terminals~\cite{footnote2}. A test of quantum mechanics through 
Bell inequalities in mesoscopic physics is very challenging and most probably it would 
be rather difficult, if not impossible, to get around all possible loopholes.
Although solid state systems are not the natural arena where to test the foundations 
of quantum mechanics, it is nevertheless very interesting to have access, manipulate and 
quantify these non-local correlations.

In this work we derive a Bell inequality for the full electron counting 
statistics and discuss its properties. The formulation we follow is based on what is 
known as the Clauser-Horne (CH) inequality~\cite{clauser74,mandel95}.
We shall show that the joint probabilities for a given number of electrons to pass 
through a mesoscopic conductor (in a given time) should satisfy, for a classical 
local theory, an inequality.

The paper is organized as follows: in the next Section 
we motivate our approach to the problem, derive the CH inequality and express the 
joint probabilities needed in the CH inequality in terms of the scattering properties 
of the mesoscopic conductor. Section \ref{results} is devoted to the discussion of the results.
We first  consider
the idealized situation where an incoming flux of fully entangled electrons is 
injected into the mesoscopic region.
Then we move on to analyze actual setups. Interacting electrons are not 
necessary to have an entangled state, we show that a three terminal normal device is enough 
to lead to violation of the CH inequality. For completeness we also consider the case 
where entanglement is produced by Andreev reflection.
In the last Section we present 
the conclusions and a brief summary of this work.

\section{CH inequality for the full counting statistics}

Electron Full Counting Statistics (FCS) refers to the probability that a given 
number of electrons has traversed, in a time $t$, a mesoscopic conductor.
In the long time limit the first and the second moment of the probability
distribution are related to the average current and noise, respectively.
The reason for which we resort to FCS for analyzing electronic entanglement
in a solid state environment resides in the fact that electrons in a conductor
are not necessarily sufficiently separated from one another for coincidence
counting to make sense, like in optical systems.
Furthermore, the measurement of single coincidence events in electronic
solid state systems does not seem realizable at present.
Zero-frequency noise accounts for long time correlations and we do not expect
it to be in general sensitive to coincidence measurements 
(see however the discussion in Ref.~\cite{samuelsson03} for the limit of small
transmission rates).
From these premises we suggest 
that FCS is a natural candidate to formulate a Bell-type inequality for electrons in 
mesoscopic conductors.
In the case where only 
two entangled electrons are injected, we find a situation similar to that 
with photons. More generally we discuss the case
where a large number of electrons have been injected.

In its original version~\cite{bell64}, the Bell inequality was derived for
dicotomic variables.
Here we consider the more general formulation due to Clauser and Horne \cite{clauser74}.
We consider the idealized setup, illustrated in Fig.\ref{Ent}, which consists of the 
following parts. On the left we place
an entangler that produces $2M$ electrons
in a spin entangled state (in Section \ref{results} two different situations 
for the implementation of the entangler are discussed).
Two conductors, characterized by some scattering matrix, connect the terminals 3 and 4
of the entangler with the exit leads 1 and 2 so to carry the two particles belonging
to each pair into two different spatially separated reservoirs.
The electron counting is performed in leads 1 and 2 for electrons with spin aligned along the 
local spin-quantization axis at angles $\theta_1$ and $\theta_2$.
Detection is realized by means of spin-selective counters, {\em i.e.}  by counting electrons
with the projection of the spin along a given local quantization-axis.
In analogy with the
optical case we say that the analyzer is not present
when the electron counting is spin-insensitive (electrons are counted
irrespective of their spin direction).

In Section \ref{derivation} we present the derivation of the CH inequality for the 
FCS and in Section \ref{scatt}
we resume, for completeness, the relation between FCS and the scattering 
matrix $S$.

\subsection{Derivation of the CH inequality}
\label{derivation}

The basic object for the  formulation of the CH inequality is
the joint probability $P(Q_1,Q_2)$ for transferring a number of $Q_1$ and  
$Q_2$ electronic charges into leads $1$ and $2$ over an observation time $t$.
We follow closely the derivation given in Ref.~\cite{mandel95}.
Our starting point is the following algebraic inequality
\begin{equation}
	-1 \le xy-xy'+x'y+x'y'-x'-y \le 0 \; 
\label{Clauser}
\end{equation}
which holds for any variable  $0 \le x,y,x',y' \le 1$.
Let us now introduce explicitly a set of hidden variables $\tau$ which take 
values in a space ${\cal T}$.
We assume that the incoming entangled electron
states are described by $\tau$ in all the details necessary to 
determine the probability distributions $P(Q_{\alpha},\tau)$ for transferring
a number of $Q_{\alpha}$ electronic charges into lead $\alpha=1,2$.
By imposing that the hidden variable theory is local, it follows that 
the joint probability can be expressed in the following form:
\begin{equation}
P(Q_1,Q_2)= \int_{\cal T} {\cal M}(\tau) P(Q_1,\tau) P(Q_2,\tau) d \tau ,
\label{jpd1}
\end{equation}
where ${\cal M}(\tau)d\tau$ defines a probability measure on the space ${\cal T}$.
The physical meaning of Eq.(\ref{jpd1}) is straightforward: it states that 
the probability distribution on lead ${\alpha}$ does not depend on the 
probability distribution on the lead ${\beta}$. 

We now introduce $P^{\theta_1,\theta_2}(Q_1,Q_2)$ as the joint 
probability for transferring $Q_1$ and $Q_2$ electronic charges when both 
analyzers are present, while $P^{\theta_1,-}(Q_1,Q_2)$ and $P^{-,\theta_2}(Q_1,Q_2)$ 
are the corresponding joint probabilities when one of the two analyzers is removed.
If the condition
\begin{equation}
P^{\theta_\alpha}(Q_\alpha,\tau) \le P(Q_\alpha,\tau) \;
\label{noen}
\end{equation}
(known as {\em no-enhancement assumption} ) is verified, it is possible to identify the 
variables appearing in Eq.(\ref{Clauser}) as follows:
\begin{eqnarray}
x&=& \frac{P^{\theta_1}(Q_1, \tau)}{P(Q_1,\tau)} ~~~ y=\frac{P^{\theta_2}
(Q_2,\tau)}{P(Q_2, \tau)} \; ,\nonumber\\
x'&=& \frac{P^{\theta'_1}(Q_1,\tau)}{P(Q_1,\tau)} ~~~ y'=\frac{P^{\theta'_2}
(Q_2, \tau)}{P(Q_2,\tau)} \; ,\label{variabili}
\end{eqnarray}
$P^{\theta}(Q_\alpha,\tau)$ being the single terminal probability distribution
in the presence of a analyzer.
Eq.(\ref{Clauser}) can then be rewritten in terms of probabilities by
multiplying
each side of the equation by $P(Q_1,\tau)P(Q_2,\tau) {\cal M}(\tau)d\tau$ 
and integrating over the space ${\cal T}$.
Finally the following inequality is obtained
\begin{eqnarray}
{\cal S}_{CH}&=&P^{\theta_1,\theta_2}(Q_1,Q_2)-P^{\theta_1,\theta'_2}(Q_1,Q_2)+
P^{\theta'_1,\theta_2}(Q_1,Q_2)+P^{\theta'_1,\theta'_2}(Q_1,Q_2)\;\nonumber\\
&-&P^{\theta'_1,-}(Q_1,Q_2)-P^{-,\theta_2}(Q_1,Q_2) \le 0 \; . 
\label{CHin}
\end{eqnarray}
Eq.(\ref{CHin})
is the CH inequality for the full counting statistics
\footnote{The CH inequality given in Eq.(\ref{CHin}) is said to be
weak in the sense that it holds only when the no-enhancement assumption, Eq.(\ref{noen}),
is satisfied.},
holding for all
values of $Q_1$ and $Q_2$ which satisfy the no-enhancement assumption.
We stress that the no-enhancement assumption, 
upon which Eq.(\ref{CHin}) is based, it is not satisfied in general
like its optical version.
The quantities that we have to compare are probability
distributions, so that Eq.(\ref{noen}) must
be checked over the whole range of $Q$.
For a fixed time $t$ and
a given mesoscopic system, hence for a given scattering matrix and incident
particle state, the no-enhancement assumption is valid only in some
range of values of $Q$.
In particular, different sets of system parameters correspond to different
such ranges.
The quantity ${\cal S}_{CH}$ in Eq.(\ref{CHin}) depends on $Q_1$ and $Q_2$ so
that the possible violation, or the extent of it, also depends on $Q_1$ and $Q_2$.
Given a certain average number $M$ of entangled pairs that have being injected in the time
$t$, one can look for the maximum violation as a function of the transmitted charges
$Q_1$ and $Q_2$.

\subsection{Scattering approach to the full counting statistics}
\label{scatt}

The joint probabilities appearing in Eq.(\ref{CHin}) can be determined once the 
scattering matrix $S$ of the mesoscopic conductor is known. The FCS in electronic 
systems was first introduced by Levitov {\it et al.} in Ref. \cite{levitov93,levitov96}
in the context of the scattering theory and later on the Keldysh Green
function method~\cite{nazarov99} to FCS was developed in Refs.\cite{nazarov01}
(for a review see Refs.~\cite{FCS}). In this paragraph we briefly describe how 
the FCS is formulated for a mesoscopic conductor in the scattering approach.
Within this framework, the transport properties of a metallic 
phase-coherent structure attached to $n$ reservoirs are determined by the matrix 
$S$ of scattering amplitudes~\cite{buttiker92}. Such amplitudes are defined through the 
scattering states describing particles propagating through the leads.
For one dimensional conductors, for example, the scattering state arising from a
unitary flux of 
particles at energy $E$ originating in the $i$-th reservoir reads
\begin{equation}
\varphi_i(x)=\frac{e^{ik_i(E)x}+r_i(E) e^{-ik_i(E)x}}{\sqrt{hv_i(E)}} ~,
\end{equation}
for the $i$-th lead, and
\begin{equation}
\varphi_j(x)=\frac{t_{ji}(E) e^{-ik_j(E)x}}{\sqrt{hv_j(E)}} ~,
\end{equation}
for the $j$-th lead, with $j\ne i$. Here $r_i(E)$ is the reflection amplitude
for particles at energy $E$, wave vector $k_i(E)$ and group velocity $v_i(E)$ 
and $t_{ji}(E)$ is the transmission amplitude from lead $i$ to lead $j$.
Note that $|r_i|^2$ is the probability for a particle to reflect back into
the $i$-th lead and $|t_{ji}|^2$ is the probability for the transmission
of a particle from lead $i$ to lead $j$. In the second quantization formalism, 
the field operator $\hat{\psi}_{j\sigma}(x,t)$ for spin $\sigma$ particles in lead 
$j$ is built from scattering states and it is defined as~\cite{lesovik}
\begin{equation}
\hat{\psi}_{j\sigma}(x,t)= \int dE ~ \frac{e^{-\frac{iEt}{\hbar}}}
{\sqrt{hv_j(E)}}
\left[ \hat{a}_{j\sigma}(E)e^{ik_j x}+\hat{\phi}_{j\sigma}(E)e^{-ik_j x}
\right] ~,
\end{equation}
where $\hat{a}_{j\sigma}(E)$ ($\hat{\phi}_{j\sigma}(E)$) is the destruction operator 
for incoming (outgoing) particles at energy $E$ with spin $\sigma$ in lead $j$.
These operators are linked by the equation
\begin{equation}
\left( \begin{array}{c} \hat{\phi}_{1\uparrow}\\ \hat{\phi}_{1\downarrow} \\
\hat{\phi}_{2\uparrow} \\ \vdots
\end{array} \right)=~S~
\left( \begin{array}{c} \hat{a}_{1\uparrow}\\ \hat{a}_{1\downarrow} \\
\hat{a}_{2\uparrow} \\ \vdots
\end{array} \right)
\label{Smat}
\end{equation}
and obey anti-commutation relations
\begin{equation}
\left\{ \hat{a}^{\dagger}_{i\sigma}(E),~\hat{a}_{j\sigma'}(E')\right\}
=\delta_{i,j} \delta_{\sigma,\sigma'} \delta(E-E') \,\,\, .
\end{equation}
In the case of two and three dimensional leads one can separate 
longitudinal and transverse particle motion. Since the transverse motion 
is quantized, the wave function relative to the plane perpendicular to 
the direction of transport is characterized by a set of quantum numbers which 
identifies the channels of the lead. Such channels are referred to as open when the 
corresponding longitudinal wave vectors are real, since they correspond to propagating 
modes. Note that the case of a single open channel corresponds to a one dimensional 
lead.

Let us now turn the attention to the probability distribution for the transfer 
of charges. Following Ref.~\cite{muzykanskii94}, within the scattering approach
the characteristic function of the probability distribution for the transfer
of particles in a structure attached to $n$ leads at a given energy $E$ can be 
written as
\begin{equation}
\chi_E(\vec{\lambda_{\uparrow}},\vec{\lambda_{\downarrow}})=
\langle \prod_{j=1,n} e^{i\lambda_{j\uparrow}
\hat{N}_I^{j\uparrow}}~  e^{\lambda_{j\downarrow} \hat{N}_I^{j\downarrow}} ~
\prod_{j=1,n}
e^{-i\lambda_{j\uparrow} \hat{N}_O^{j\uparrow}}  e^{{-i\lambda_{j\downarrow}
\hat{N}_O^{j\downarrow}}} \rangle ~,
\label{chiE}
\end{equation}
where the brackets $\langle ... \rangle$ stand for the quantum statistical
average over the thermal distributions in the leads.
Assuming a single channel per lead, $\hat{N}_{I(O)}^{j\sigma}$ is the number
operator for incoming (outgoing) particles with spin $\sigma$ in lead $j$
and $\vec{\lambda_{\uparrow}}$, $\vec{\lambda_{\downarrow}}$ are vectors
of $n$ real numbers, one for each open channel. In terms of incoming (outgoing) creation 
operator the number operators can be expressed as follows
\begin{equation}
\hat{N}_I^{j\sigma}= \hat{a}_{j\sigma}^{\dagger}\hat{a}_{j\sigma}; \qquad
\hat{N}_O^{j\sigma}=\hat{\phi}_{j\sigma}^{\dagger}\hat{\phi}_{j\sigma} \;\; .
\end{equation}
Eq.(\ref{chiE}) can also be recasted in the form~\cite{levitov93}:
\begin{equation}
\chi_E(\vec{\lambda_{\uparrow}},\vec{\lambda_{\downarrow}})=
\det (\mathbb{I}-n_E+n_E~ S^\dagger \Lambda^\dagger~ S~ \Lambda) ,
\label{chiEdet}
\end{equation}
where $\mathbb{I}$ is the unit matrix, $n_E$ is the diagonal matrix of Fermi
distribution functions $f_j(E)$
for particles in the reservoir $j$ and defined as
$(n_E)_{j\sigma,j\sigma}=f_j(E)$, whereas $\Lambda$ is a diagonal matrix
defined as: $(\Lambda)_{j\sigma,j\sigma}=\exp (i\lambda_{j\sigma})$.
For long measurement times $t$ the total characteristic
function $\chi$ is the product of contributions from different energies, so that
\begin{equation}
\chi(\vec{\lambda_{\uparrow}},\vec{\lambda_{\downarrow}})=e^{\frac{t}{h}\int
dE~\log{\chi_E(\vec{\lambda_{\uparrow}},\vec{\lambda_{\downarrow}})}} \; .
\label{chi}
\end{equation}
At zero temperature, the statistical average over the Fermi distribution
function in Eq.(\ref{chiE}) simplifies to the expectation value calculated on the state
$\ket{\psi}$ containing two electrons of both spin species for each channel of a given lead
up to the energy corresponding to the chemical potential of such lead.
Furthermore, in the limit of a small bias voltage $V$ applied between the reservoirs,
the argument of the integral is energy-independent so that
Eq.(\ref{chi}) can be approximated to
\begin{equation}
\chi(\vec{\lambda_{\uparrow}},\vec{\lambda_{\downarrow}})\simeq
\left[ \chi_0(\vec{\lambda_{\uparrow}},\vec{\lambda_{\downarrow}})\right]^M
\label{approx}
\end{equation}
where only the zero-energy characteristic function appears and $M=eVt/h$ is
the average number of injected particles.
The joint probability distribution for transferring
$Q_{1\sigma}$ spin-$\sigma$ electrons in lead 1, $Q_{2\sigma}$
spin-$\sigma$ electrons in lead 2, etc. is related to the characteristic function by 
the relation (we assume that no polarizers are present):
\begin{equation}
P(Q_{1\uparrow},Q_{1\downarrow},Q_{2\uparrow},\ldots)=\frac{1}{(2\pi)^{2n}}
\int_{-\pi}^{+\pi} d\lambda_{1\uparrow} d\lambda_{1\downarrow}
d\lambda_{2\uparrow}
\ldots ~\chi(
\vec{\lambda_{\uparrow}},\vec{\lambda_{\downarrow}}
) ~e^{i\vec{\lambda_{\uparrow}}\cdot\vec{Q_{\uparrow}}}
~e^{i\vec{\lambda_{\downarrow}}\cdot\vec{Q_{\downarrow}}} ~.
\label{counting}
\end{equation}

In the rest of the paper we will consider systems where only two counting
terminals are present.
In particular, while the counting terminals are kept at the lowest chemical potential,
all other terminals are biased at chemical potential $eV$.
For later convenience, we write down the most general expression
for the characteristic function when spin-$\sigma$ electrons are counted in lead
1 and spin-$\sigma'$ electrons are counted in lead 2:
\begin{eqnarray}
\chi_E (\lambda_{1\sigma},\lambda_{2\sigma'})=1+
\left( e^{-i\lambda_{1\sigma}} -1 \right) \langle \hat{N}_{\text{O}}^{1\sigma} \rangle +
\left( e^{-i\lambda_{2\sigma'}} -1 \right) \langle \hat{N}_{\text{O}}^{2\sigma'} \rangle +
\nonumber\\ +
\left( e^{-i\lambda_{1\sigma}} -1 \right) \left( e^{-i\lambda_{2\sigma'}} -1 \right)
\langle \hat{N}_{\text{O}}^{1\sigma} \hat{N}_{\text{O}}^{2\sigma'} \rangle ,
\label{chi_s}
\end{eqnarray}
in the relevant energy range $0<E<eV$.
The parameters $\lambda$ corresponding to all others terminals are set to zero.

Using Eqs.~(\ref{approx}), (\ref{counting}) and (\ref{chi_s}), at zero temperature,
one can calculate the single terminal probability distribution:
\begin{equation}
P(Q_{1\sigma})=\left( \begin{array}{c}M\\Q_{1\sigma} \end{array}\right)
\left[ 1- \bra{\psi} \hat{N}_{\text{O}}^{1\sigma} \ket{\psi}\right]^{M-Q_{1\sigma}} \bra{\psi} 
\hat{N}_{\text{O}}^{1\sigma} \ket{\psi}^{Q_{1\sigma}} 
\label{stp}
\end{equation}
and the joint probability distribution:
\begin{eqnarray}\nonumber
P (Q_{1\sigma},Q_{2\sigma'})=\sum_{k=\textrm{Max}[M-Q_{1\sigma},M-Q_{2\sigma'}]}^{(M-Q_{1\sigma})+(M-Q_{2\sigma'})}
A^{2M-Q_{1\sigma}-Q_{2\sigma'}-k} ~B^{Q_{1\sigma}-M+k} ~C^{k-M+Q_{2\sigma'}}\times \\
\times \bra{\psi} \hat{N}_{\text{O}}^{1\sigma} \hat{N}_{\text{O}}^{2\sigma'} \ket{\psi}^{M-k}  f(M,Q_{1\sigma},Q_{2\sigma'},k)
\label{FCSt}
\end{eqnarray}
where
$
A=1- \bra{\psi}\hat{N}_{\text{O}}^{1\sigma} \ket{\psi} -\bra{\psi}
\hat{N}_{\text{O}}^{2\sigma'} \ket{\psi}
+ \bra{\psi} \hat{N}_{\text{O}}^{1\sigma} \hat{N}_{\text{O}}^{2\sigma'} \ket{\psi}
$,
$
B=\bra{\psi} \hat{N}_{\text{O}}^{1\sigma} (1- \hat{N}_{\text{O}}^{2\sigma'})\ket{\psi}
$, 
$
C=\bra{\psi} (1- \hat{N}_{\text{O}}^{1\sigma})\hat{N}_{\text{O}}^{2\sigma'} \ket{\psi} 
$
and
$
f(M,Q_{1\sigma},Q_{2\sigma'},k) =M!/[(k-M+Q_{2\sigma'})!(2M-k-Q_{1\sigma}-Q_{2\sigma'})!] 
$.
In doing so we have written the expressions for the probability distributions
in terms of the expectation values of ``outgoing'' number operators.
For $Q_{1\sigma}=Q_{2\sigma'}=M$, Eq.(\ref{FCSt}) reduces to
\begin{equation}
P(Q_{1\sigma}=M,Q_{2\sigma'}=M)=\bra{\psi} \hat{N}_{\text{O}}^{1\sigma} \hat{N}_{\text{O}}^{2\sigma'} \ket{\psi}^M .
\end{equation}

When both spin species are counted in one of the terminals
the characteristic function is different from the one given in Eq.(\ref{chi_s}).
In particular, the characteristic function for counting both spins in terminal 1 reads:
\begin{eqnarray}\nonumber
\chi_E (\lambda_1,\lambda_{2\sigma'})=1+
\left( e^{-i\lambda_1} -1 \right) \langle \left( \hat{N}_{\text{O}}^{1\uparrow} +
\hat{N}_{\text{O}}^{1\downarrow}\right) \rangle +
\left( e^{-i\lambda_{2\sigma'}} -1 \right) \langle \hat{N}_{\text{O}}^{2\sigma'} \rangle +
\\\nonumber+
\left( e^{-i\lambda_1} -1 \right) \left( e^{-i\lambda_{2\sigma'}} -1 \right)
\langle \left( \hat{N}_{\text{O}}^{1\uparrow}+
\hat{N}_{\text{O}}^{1\downarrow}\right) \hat{N}_{\text{O}}^{2\sigma'} \rangle +
\left( e^{-i\lambda_1} -1 \right)^2
\langle \hat{N}_{\text{O}}^{1\uparrow} \hat{N}_{\text{O}}^{1\downarrow} \rangle +
\\
\left( e^{-i\lambda_1} -1 \right)^2 \left( e^{-i\lambda_{2\sigma'}} -1 \right)
\langle \hat{N}_{\text{O}}^{1\uparrow} \hat{N}_{\text{O}}^{1\downarrow} 
\hat{N}_{\text{O}}^{2\sigma'} \rangle .
\label{chi_s1}
\end{eqnarray}
where we have set $\lambda_{1\uparrow}=\lambda_{1\downarrow}\equiv \lambda_1$.
The expression for the joint probability distribution is in general complicated,
as one can see in Appendix \ref{PR} where such expressions for different systems
are reported.

\section{Results}
\label{results}

The inequality presented in Eq.(\ref{CHin}) can be tested in various multi-terminal 
mesoscopic conductors. In this Section we present several geometries that can be 
experimentally realized. In order to get acquainted with the informations that can 
be retrieved from  Eq.(\ref{CHin}) we start from an ideal case in which the entangled 
pair is generated by some {\em entangler} in the same spirit as in the works of 
Refs.~\cite{burkard00,taddei02}.
In Section \ref{ns} we shall demonstrate that a normal beam splitter in the absence of interaction
is enough to generate entangled pairs of electrons, therefore constituting a 
simple realization of an entangler.
For comparison we also analyze the role of superconductivity in creating spin singlets.

\subsection{Entangled electrons}
\label{ee}

\begin{figure}
\begin{center}
\epsfig{figure=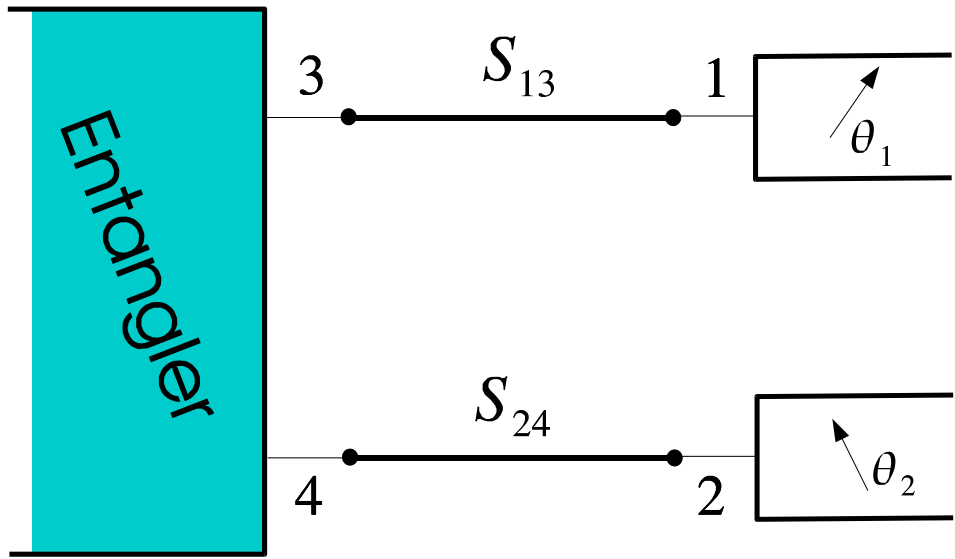,width=0.8 \textwidth}
\end{center}
\caption{Idealized setup for testing the CH inequality for electrons in a
solid state environment. It consists of two parts: an entangler
(shaded block) that produces pairs of spin entangled electrons
exiting from terminals 3 and 4. These terminals are connected to leads 1 and 2
through two conductors described by scattering
matrices $S_{13}$ and $S_{24}$.
Electron counting is performed in leads 1 and 2 
along the local spin-quantization
axis oriented at angles $\theta_1$ and $\theta_2$.}
\label{Ent}
\end{figure}

In the setup depicted in Fig.~\ref{Ent}
we assume the existence of an entangler that produces electron pairs in the Bell state
\begin{equation}
\ket{\psi}=\frac{1}{\sqrt{2}} 
\left [a^{\dagger}_{3\uparrow}(E) a^{\dagger}_{4\downarrow}(E) \pm  a^{\dagger}_{3\downarrow}(E)
a^{\dagger}_{4\uparrow} (E)\right ]
\ket{0} ,
\label{st}
\end{equation}
of spin triplet (upper sign) or spin singlet (lower sign) in the energy range $0<E<eV$.
These electrons propagate through the conductors which connect terminals 3 and 4 with
leads 1 and 2, as though terminals 3 and 4 were kept at a potential $eV$
with respect to 1 and 2.
Our aim is to test the violation of the 
CH inequality given in Eq.(\ref{CHin}) for such maximally entangled 
states.

When the angles $\theta_1$ and $\theta_2$ are parallel to each other,
the scattering matrix of the two conductors,
in the absence of spin mixing processes, can be written as:
\begin{equation}
S= \left( \begin{array}{cc}
\hat{S}_{13} & 0 \\
0 & \hat{S}_{24}
\end{array} \right)\;
\label{S-ent}
\end{equation}
where
\begin{equation}
\hat{S}_{13}=\left (\begin{array}{cc}
\check{r}_{3} & \check{t}_{31} \\
\check{t}_{13} & \check{r}_{1}
\end{array} \right )= \left (\begin{array}{cccc}
r_{3 \uparrow} & 0 & t_{31 \uparrow} & 0 \\
0 &  r_{3 \downarrow} & 0 &  t_{31 \downarrow} \\
t_{13 \uparrow} & 0 & r_{1 \uparrow} & 0 \\
0 &  t_{13 \downarrow} & 0 &  r_{1 \downarrow} 
\end{array} \right ) .
\end{equation}
Here $r_{j\sigma}$ ($t_{ij \sigma}$) is the probability amplitude for an incoming particle with
spin $\sigma$  from lead $j$ to be reflected (transmitted in lead $i$).
For a normal-metallic wire we set
$t_{ij \uparrow}=t_{ij \downarrow}=\sqrt{T}$, $t_{ji \uparrow}=t_{ji \downarrow}=-\sqrt{T}$
and $r_{j\uparrow}=r_{j\downarrow}=\sqrt{1-T}$, where $T$ is the transmission probability.
The expression for $\hat{S}_{24}$ is written analogously.
For simplicity we will assume that $\hat{S}_{13}$ and $\hat{S}_{24}$ are equal.
The general scattering matrix relative to non-collinear angles is obtained from
$S$ by rotating the spin quantization axis independently in the two conductors
(note that this is possible because the two wires are decoupled).
The ``rotated'' S-matrix is obtained \cite{brataas01} by the transformation
$S_{\theta_1,\theta_2}={\cal U} S {\cal U}^{\dagger}$, where ${\cal U}$
is the rotation matrix given by:
\begin{equation}
{\cal U}=\left( \begin{array}{cccc} 
U_{\theta_1} & 0 &0 & 0\\
0&\mathbb{I}&0 & 0\\
0 & 0 & U_{\theta_2} & 0 \\
0&0&0& \mathbb{I}\end{array} \right)
\label{tra}
\end{equation}
where
\begin{equation}
U_{\theta}=\left( \begin{array}{cc} \cos{\theta\over 2}&\sin{\theta\over 2}\\
-\sin{\theta\over 2}&\cos{\theta\over 2}
\end{array} \right) .
\label{uteta}
\end{equation}
The probability distributions are now given by the expressions in 
Eq.~(\ref{stp}) and Eq.~(\ref{FCSt})
where the state $\ket{\psi}$ is given by Eq.~(\ref{st}).
In the case where both analyzers are present we set $\sigma=\sigma'=\uparrow$.
The probability distribution when one of the analyzers is removed also possesses
the structure of Eq.(\ref{FCSt}) since, in this case, the correlators 
$\langle \hat{N}_{\text{O}}^{1\uparrow} \hat{N}_{\text{O}}^{1\downarrow} \rangle$ and
$\langle \hat{N}_{\text{O}}^{1\uparrow} \hat{N}_{\text{O}}^{1\downarrow} 
\hat{N}_{\text{O}}^{2\uparrow} \rangle$ appearing in Eq.(\ref{chi_s1})
vanish.
In particular when, for example, the upper analyzer in Fig. \ref{Ent} is removed we
need to replace
$\hat{N}_{\text{O}}^{1\sigma}$ with $\hat{N}_{\text{O}}^{1\uparrow}+\hat{N}_{\text{O}}^{1\uparrow}$
and $\hat{N}_{\text{O}}^{2\sigma'}$ with $\hat{N}_{\text{O}}^{2\uparrow}$.
For the other correlators one gets:
\begin{equation}
\bra{\psi}\hat{N}_{\text{O}}^{1\uparrow}\ket{\psi}=
\bra{\psi}\hat{N}_{\text{O}}^{2\uparrow}\ket{\psi}=\frac{T}{2} ,
\label{27}
\end{equation}
\begin{equation}
\bra{\psi}\hat{N}_{\text{O}}^{1\downarrow}\ket{\psi}=\frac{T}{2} ,
\label{28}
\end{equation}
\begin{equation}
\bra{\psi}\hat{N}_{\text{O}}^{1\uparrow}\hat{N}_{\text{O}}^{2\uparrow} \ket{\psi}=
\frac{T^2}{2} \sin^2 \left( \frac{\theta_1\pm\theta_2}{2}\right)
\end{equation}
and
\begin{equation}
\bra{\psi}\hat{N}_{\text{O}}^{1\downarrow}\hat{N}_{\text{O}}^{2\uparrow} \ket{\psi}=
\frac{T^2}{2} \cos^2 \left( \frac{\theta_1\pm\theta_2}{2}\right) .
\end{equation}

For the single terminal probability distributions in leads $i=1,2$ we get,
in the presence and in the absence of an analyzer, respectively,
\begin{eqnarray}
P^{\theta_i}(Q_i)&=&\left( \begin{array}{c}M\\Q_i \end{array}\right) 
\left (\frac{T}{2} \right )^{Q_i} \left (1-\frac{T}{2} \right )^{M-Q_i} \label{prob00}\\
P(Q_i)&=&\left( \begin{array}{c}M\\Q_i \end{array}\right)
\left (T \right )^{Q_i} \left (1-T \right )^{M-Q_i} ,
\label{prob0}
\end{eqnarray}
so that the no-enhancement assumption reads:
\begin{equation}
\left (1-\frac{T}{2} \right )^{(M-Q_i)} \left (\frac{1}{2} \right )^{Q_i} \le
(1-T)^{(M-Q_i)} ~~~~~~~i=1,2\; .
\label{noen1}
\end{equation}
Note that the probabilities in Eqs.~(\ref{prob00}) and (\ref{prob0}) do not depend
on the angles $\theta_1$ and $\theta_2$ because the expectation values in Eqs.~(\ref{27})
and (\ref{28}) are invariant under spin rotation.
As a consequence, the effect of the analyzer is equivalent to a reduction of the
transmission probability $T$ by a factor of 2, resulting in a shift of the
maximum of the distribution.
From Eq.(\ref{noen1}) it follows that, for a given number $M=eVt/h$ of entangled pairs generated by
the entangler, the no enhancement assumption can  be verified only for certain values of
$T$ and of $Q_i$.
This makes clear that the CH inequality of Eq.(\ref{CHin}) can be tested for
violation only for appropriate values of $M$, $T$ and $Q_1$ or $Q_2$.
For example, for a given observation time $t$ ({\em i.e.} a given $M$)
and a given value of $Q$, CH inequality can
be tested only for transmission $T$ less than a maximum value given by the 
expression 
\begin{equation}
T_{\text{max}}=\frac{2^{\frac{Q_i}{M-Q_i}}-1}{2^{\frac{Q_i}{M-Q_i}}-\frac{1}{2}} .
\label{condition}
\end{equation}
At the edge of the distribution ($Q_i=M$)
the no-enhancement assumption is satisfied for every $T$.
The window of allowed $Q_i$ values where the no-enhancement assumption is satisfied
gets wider on approaching the tunneling limit. 
For large $M$, $T_{\text{max}}\simeq  2 (\log2) \frac{Q_i}{M}$.
The previous inequality can be also interpreted as a limit for the allowed 
measuring time given a setup at disposal. 
Alternatively, given a certain transmission, the no-enhancement assumption is verified
for points of the distribution such that:
\begin{equation}
\frac{Q_i}{M}\geq \frac{\log\frac{1-T/2}{1-T}}{\log 2+\log\frac{1-T/2}{1-T}} .
\label{ultimo}
\end{equation}

The various probabilities needed to define ${\cal S}_{CH}$ are
collected in Appendix \ref{PR}.
However, it is useful to note here that the
joint probabilities with a single analyzer are factorized:
\begin{eqnarray}\nonumber
P^{\theta_1,-}(Q_1,Q_2)=P^{\theta_1}(Q_1) P(Q_2)\\
P^{-,\theta_2}(Q_1,Q_2)=P(Q_1) P^{\theta_2}(Q_2) ,
\end{eqnarray}
while joint probabilities with two analyzers are not factorized.
Furthermore, all such probabilities have a common factor, $T^{Q_1+Q_2}/2^M$,
which leads to an exponential suppression for large $M$ and $Q_1+Q_2$.
We shall address the question of whether this also produces a suppression of
${\cal S}_{CH}$ in case of violation.

\begin{figure}
\begin{center}
\epsfig{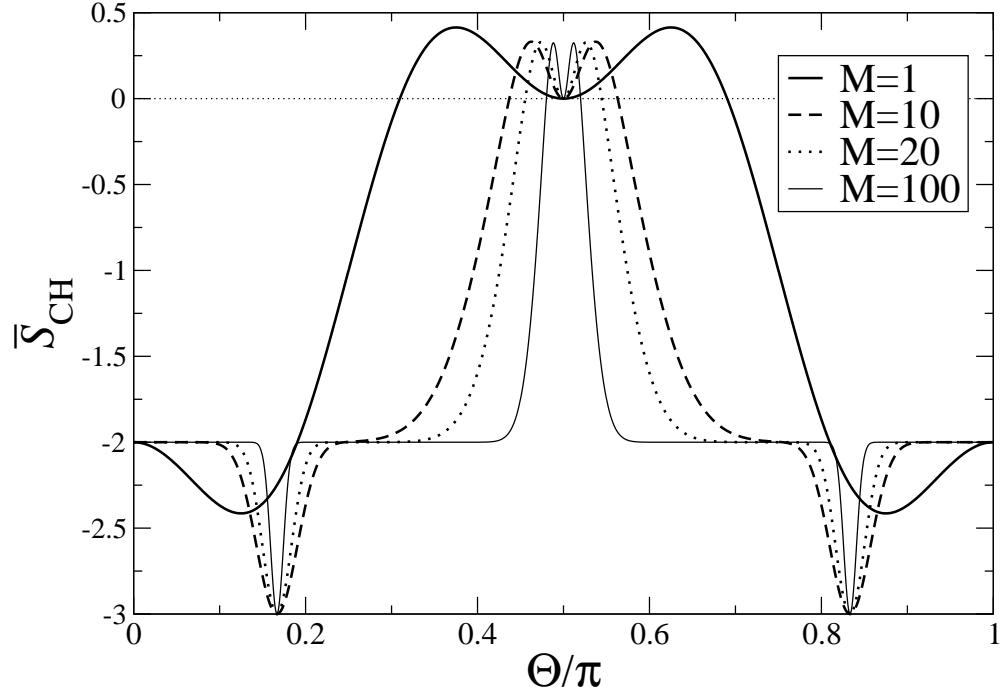}
\end{center}
\caption{The quantity $\overline{{\cal S}}_{CH}={\cal S}_{CH}/(T^{2M}/2^M)$ is plotted as a
function
of the angle $\Theta$ for different numbers $M$ of injected entangled pairs by the entangler.
The range of angles relative to positive values shrinks with increasing $M$,
while the value of the maximum slightly decreases.}
\label{Srid:Q=M}
\end{figure}

Let us now analyze the possibility of violation of the CH inequality for different values of $Q_1$ and $Q_2$.
First consider the situation where the entangler emits a single entangled pair of
electrons in which case $P^{\theta_1,\theta_2}(1,1)=\bra{\psi}
\hat{N}_{\text{O}}^{1\uparrow} \hat{N}_{\text{O}}^{2\uparrow}\ket{\psi}$,
$P^{-,\theta_2}(1,1)=\bra{\psi}(
\hat{N}_{\text{O}}^{1\uparrow}+\hat{N}_{\text{O}}^{1\downarrow})
\hat{N}_{\text{O}}^{2\uparrow}\ket{\psi}$ and
$P^{\theta_1,-}(1,1)=\bra{\psi}
\hat{N}_{\text{O}}^{1\uparrow} (\hat{N}_{\text{O}}^{2\uparrow}+
\hat{N}_{\text{O}}^{2\uparrow})\ket{\psi}$.
We find that the
CH inequality is maximally violated for the following choice of angles:
$\theta_2-\theta_1=\theta'_2-\theta'_1=3\pi /4$.
More precisely we obtain:
\begin{equation}
{\cal S}_{CH}= T^2 \frac{\sqrt{2}-1}{2}
\end{equation}
which is equal to the result obtain for an entangled pair of photons \cite{mandel95},
where $T$ plays the role of the quantum efficiency of the photon detectors.
In the more general case of $Q_1=Q_2=M$, for $M\gg 1$, we have
\begin{eqnarray}
P^{\theta_1,\theta_2}(M,M)&=&\frac{T^{2M}}{2^M} 
\left [\sin^2 \left ( \frac{\theta_1 \pm \theta_2}{2} \right ) \right ]^M \; \nonumber \\
P^{\theta_1,-}(M,M)&=& P^{-,\theta_2}(M,M)=\frac{T^{2 M}}{2^M}  \;\label{prob}
\end{eqnarray}
so that the no-enhancement assumption is always satisfied and the quantity
${\cal S}_{CH}$ can be easily evaluated:
\begin{equation}
{\cal S}_{CH}=\frac{T^{2M}}{2^M} \left[ \sin^{2M}\frac{\theta_1\pm \theta_2}{2}-
\sin^{2M}\frac{\theta_1\pm \theta'_2}{2} +
\sin^{2M}\frac{\theta'_1\pm \theta_2}{2} +
\sin^{2M}\frac{\theta'_1\pm \theta'_2}{2} -2
\right] .
\end{equation}
The rotational invariance makes $P^{\theta_1,-}$ and $P^{-,\theta_2}$
independent of angles, and $P^{\theta_1,\theta_2}$ dependent on the
angles through $\frac{\theta_1\pm \theta_2}{2}$.
This allows us, without loss of generality, to define
an angle $\Theta$ such that $2\Theta=\theta_1\pm \theta_2=\theta'_1\pm \theta_2=
\theta'_1\pm \theta'_2=(\theta_1\pm \theta'_2)/3$.
As a result Eq.(\ref{CHin}) takes the form:
\begin{equation}
{\cal S}_{CH}=3P_{1,2}^{\Theta}(Q_1,Q_2)-P_{1,2}^{3\Theta}(Q_1,Q_2)-P_{1,-}(Q_1,Q_2)-P_{-,2}(Q_1,Q_2)
\le 0
\end{equation}
where $P_{1,2}^{\Theta}=P^{\theta_1,\theta_2}$ and $P_{1,-}=P^{\theta_1,-}$.
It is useful to
define the reduced quantity $\overline{{\cal S}}_{CH}={\cal S}_{CH}/(T^{2M}/2^M)$ which
is plotted in Fig. \ref{Srid:Q=M} as a function of $\Theta$ for different
values of $M$ (note that since $P^{\theta_1,-}(M,M)=(T^{2M}/2^M)$, $\overline{{\cal S}}_{CH}$
is nothing but ${\cal S}_{CH}/P^{\theta_1,-}(M,M)$).
The violation occurs for every value of $M$ in a range of angles around $\Theta=\pi/2$
(note that ${\cal S}_{CH}$ is symmetric with respect to $\pi/2$).
The range of angles for which $\overline{{\cal S}}_{CH}$ is positive shrinks
with increasing $M$, while the maximum value of $\overline{{\cal S}}_{CH}$ decreases very weakly
with $M$ (more precisely, $\overline{{\cal S}}_{CH}^{\text{max}}\propto 1/M$).
This means that
the effect of the factor $T^{2M}/2^M$ on the value of ${\cal S}_{CH}$ is
exponentially strong, making the violation of the CH inequality exponentially difficult
to detect for large $M$ and $Q_1=Q_2=M$.
The weakening of the violation is mainly due to
the suppression of the joint probabilities.
As we shall show later, by optimizing all the parameters it is yet possible
to eliminate this exponential suppression.

\begin{figure}
\begin{center}
\epsfig{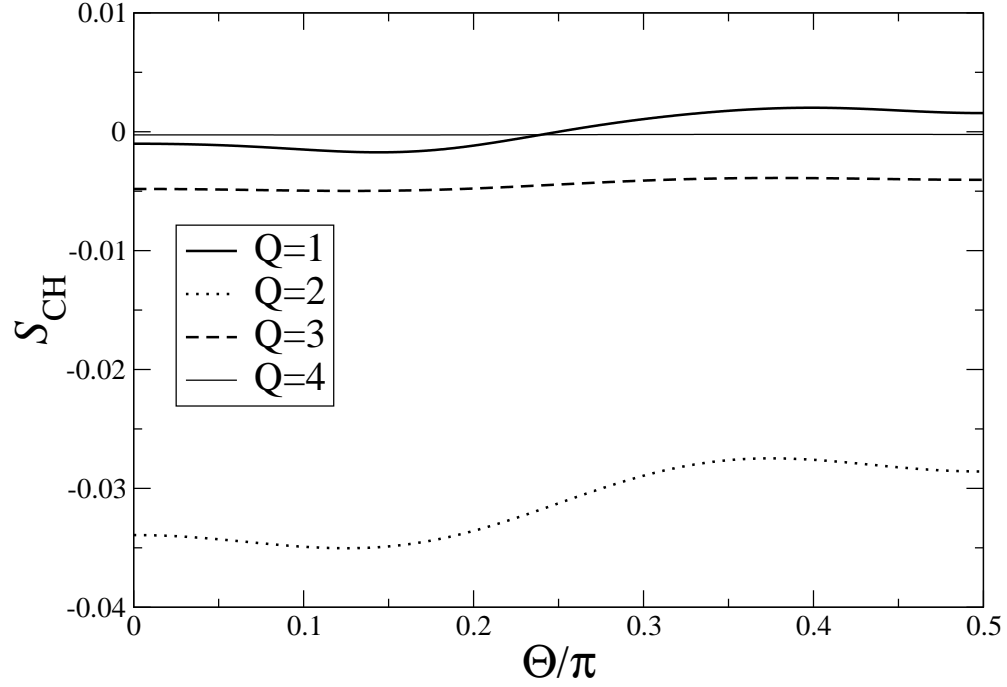}
\end{center}
\caption{The quantity ${\cal S}_{CH}$ is plotted as a function of the angle $\Theta$
for $M=20$ and $T=0.06917$, which corresponds to the highest value allowed by
the no-enhancement assumption for $Q=1$.
The curves are relative to different values of $Q=[1,4]$. Note that for $Q\ge 4$ the
variation of ${\cal S}_{CH}$ over the whole range of $\Theta$ is small on the
scale of the plot.
Violations are found only for $Q=1$ and $Q=20$.
}
\label{S:Q<M}
\end{figure}

\begin{figure}
\begin{center}
\epsfig{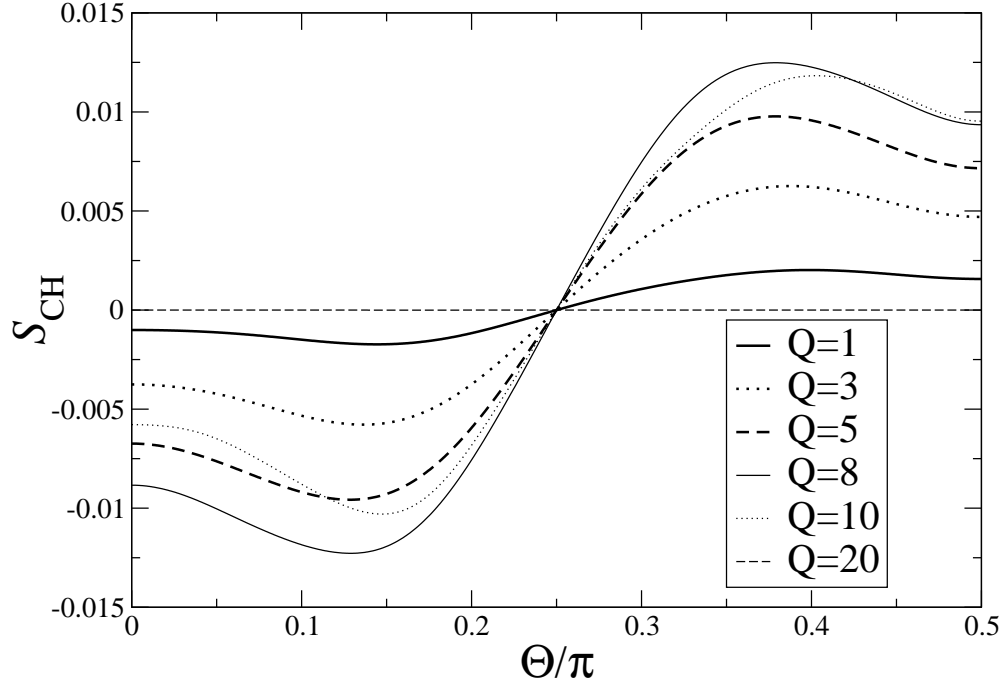}
\end{center}
\caption{The quantity ${\cal S}_{CH}$ is plotted as a function of the angle $\Theta$
for $M=20$ and $T$ set to the highest value allowed by the no-enhancement assumption,
different from each $Q$.
The curves are relative to different values of $Q=[1,20]$.
The maximum of ${\cal S}_{CH}$ increases with $Q$ reaching its largest value for $Q=8$ and
decreasing for $Q>8$.
Note that the variation of ${\cal S}_{CH}$ with $\Theta$ for $Q=20$ it is not appreciable
on this scale.
}
\label{S1:Q<M}
\end{figure}

Let us now consider the violation of the CH inequality as a function of
the transmitted charges. We notice that the CH inequality is not violated for the 
off-diagonal terms of the distributions (when $Q_1 \neq Q_2$), meaning that one
really needs to look at ``coincidences''.
Therefore we discuss the case 
$Q_1=Q_2\equiv Q<M$ (remember that the no-enhancement assumption is satisfied only
for $T\leq T_{\text{max}}(Q)$).
In Fig. \ref{S:Q<M} we plot the quantity ${\cal S}_{CH}$ for $M=20$ as a function
of $\Theta$ and different values of $Q$.
The transmission $T$ is fixed at the highest allowed value by the no-enhancement assumption,
which corresponds to the smallest $Q$ considered $T_{\text{max}}(Q=1)=0.06917$.
Fig. \ref{S:Q<M} shows that the largest positive value of ${\cal S}_{CH}$ and the
widest range of angles
corresponding to positive ${\cal S}_{CH}$ occur for $Q=1$, {\em i.e.} for a joint probability
relative to the detection of a single pair.
One should not conclude that, in order to detect the violation of the CH inequality,
only very small values of the transmitted charge should be taken. We have in fact
considered $T=T_{\text{max}}$ relative
to $Q=1$ and the maximum violation, for given $M$ and $Q$, always occurs
at $T=T_{\text{max}}$.
In order to get the largest violation of the CH inequality
at a given $M$ and $Q$ one could, in principle, choose the highest allowed value
of $T$ for each value of $Q$ ($T=T_{\text{max}}(Q)$).
We show in Fig. \ref{S1:Q<M} the corresponding plot, to be compared with Fig. \ref{S:Q<M}.
For every $Q<M$ the violation occurs in the same range of angles,
namely $\pi/4 \le\Theta\le \pi/2$, 
because of the following properties of the joint probability distributions:
$P^{\Theta}_{1,2}(Q_1,Q_2)=P^{3\Theta}_{1,2}(Q_1,Q_2)=P_{1,-}(Q_1,Q_2)$ for $\Theta=\pi/4$.
This implies that ${\cal S}_{CH}(\Theta=\pi/4)=0$,
and $P^{\Theta}_{1,2}(Q_1,Q_2)\geq P^{3\Theta}_{1,2}(Q_1,Q_2),P^{\Theta}_{1,-}(Q_1,Q_2),
P^{\Theta}_{-,2}(Q_1,Q_2)$ for $\pi/4 \le\Theta\le \pi/2$.
Furthermore, in this specific case of $M=20$, we find that
the maximum values of $S$ occurs at $Q=8$.

\begin{figure}
\begin{center}
\epsfig{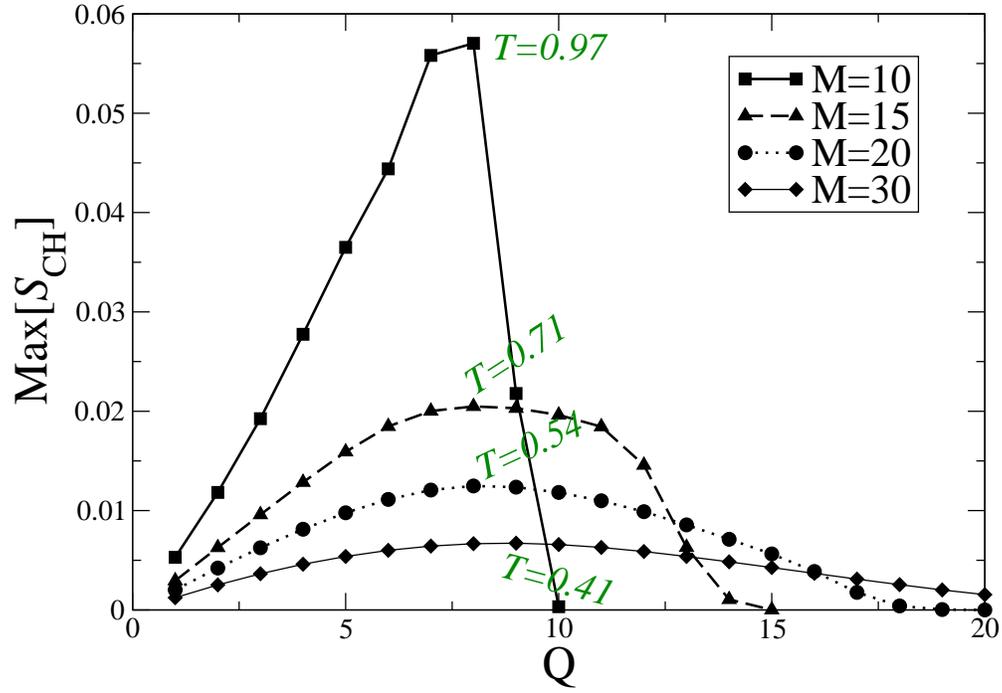}
\end{center}
\caption{The maximum value of the quantity ${\cal S}_{CH}$, evaluated over angles $\Theta$
and transmission probabilities $T$, is plotted as a function of $Q$.
The curves are relative to different values of $M$ ranging from 10 to 30.
For points corresponding to the maximum of the curves we indicate the corresponding value
of transmission $T$.
}
\label{Smax:Q<M}
\end{figure}

In Fig.~\ref{Smax:Q<M} we plot the maximum value of $S$, with respect to $\Theta$ and $T$,
as a function of $Q$ for different values of $M$.
Several observations are in order. For increasing $M$, the position of the maximum,
$Q_{\text{max}}$ is very weakly dependent on $M$.
Remarkably, the value of the maximum of the curves
does not decreases exponentially, but rather as $1/M^2$.
Despite the exponential suppression of the joint probability with $M$, the extent of the
maximal violation scales with $M$ much slowly (polynomially).

\begin{figure}
\begin{center}
\epsfig{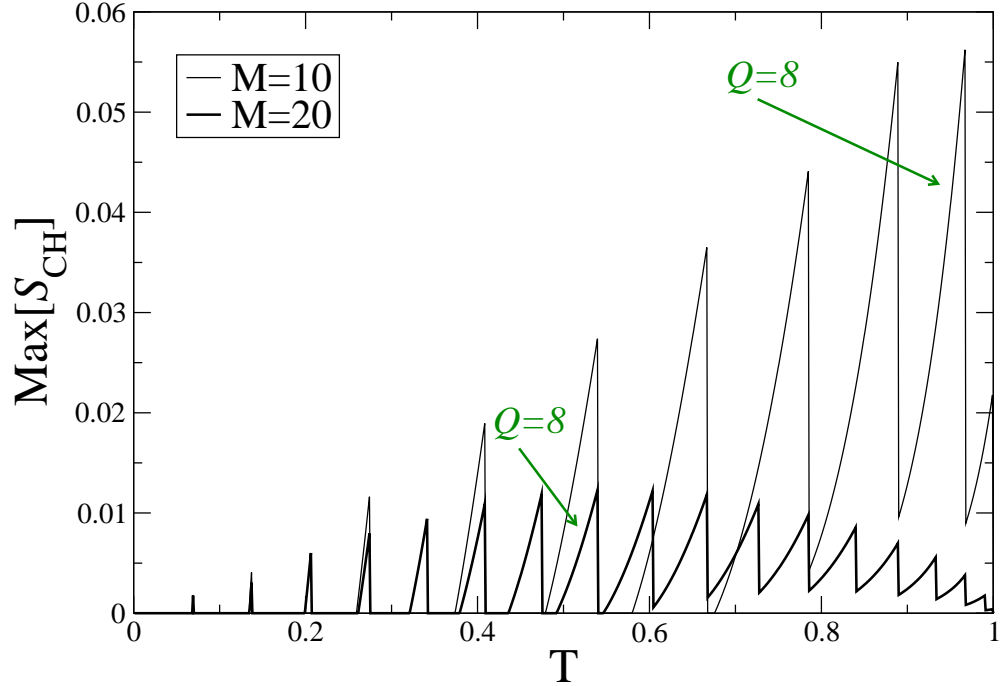}
\end{center}
\caption{The maximum value of the quantity ${\cal S}_{CH}$, evaluated over angles $\Theta$
and number of counted electrons $Q$, is plotted as a function of $T$.
Both curves, relative to $M=10$ and $M=20$, exhibit discontinuities which correspond to
an increase of the value of $Q$ by one.
This increase is imposed by the no-enhancement assumption, Eq.~(\ref{ultimo}), which depends
on the value of $T$.
We indicate the value of $Q$ which corresponds to the largest violation.
}
\label{Smax:T}
\end{figure}

It may be useful to look at the same situation from a different perspective.
Given a certain transmission $T$ ({\em i.e.} fixing the transport properties of the conductors)
we want to find when the CH inequality is maximally violated.
For a given observation time $t$, the no-enhancement assumption Eq.~(\ref{condition})
imposes a minimum value for $Q$.
In Fig.~\ref{Smax:T} we plot the quantity ${\cal S}_{CH}$, maximized over the angle $\Theta$
and $Q$, as a function of $T$ for different $M$.
The curves are piecewise increasing function of $T$, where the discontinuities correspond to an
increase of the value of $Q$ by one imposed by the no-enhancement assumption.
More precisely,
when $T$ is increased above a threshold for which Eq.~(\ref{condition}) is not satisfied,
one needs to increase $Q$ by one unit in order for this condition to be recovered.
The result of this is a jump in the values of the probabilities that leads to
a discontinuity of ${\cal S}_{CH}$.
Fig.~\ref{Smax:T} allows to choose the best values of $M$ and $Q$ to get the maximum
violation.

If the entangler is substituted with a source that emits factorized states, 
the CH inequality given in Eq.(\ref{CHin}) is never 
violated. In this case, in contrast to Eq.(\ref{st}), the state emitted by the source reads:
$
\ket{\psi}=a^{\dagger}_{3\uparrow} a^{\dagger}_{4\uparrow} \ket{0} .
$
All the previous calculations can be repeated and we find, as expected, that 
the characteristic functions factorizes, so that the two terminal joint probability 
distributions are given by the product of the single terminal probability distributions.

\subsection{Normal beam splitter}
\label{ns}

\begin{figure}
\begin{center}
\epsfig{figure=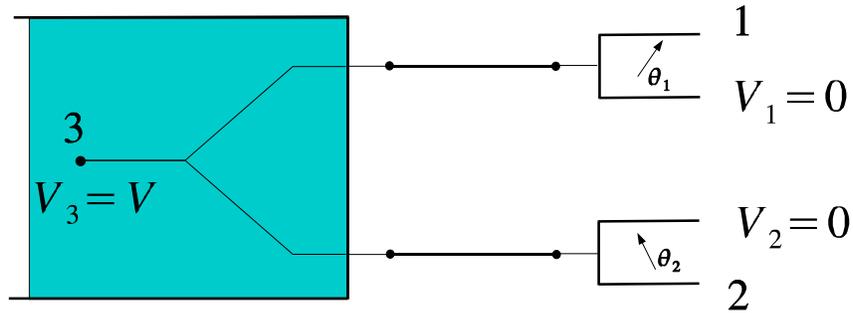,width=0.7 \textwidth}
\end{center}
\caption{Setup of a realistic system consisting of a normal beam splitter (shaded
region) for testing the CH inequality.
Bold lines represent two conductors of unit transmission probability.
A bias voltage equal to $eV$ is set between terminals 3 and 1 and terminals 3 and 2.
}
\label{Stub}
\end{figure}

We are now ready to analyze realistic structures by replacing the shaded
block in Fig.~\ref{Ent} (which represents the entangler) with a certain system, and discuss
the CH inequality along the lines of Section \ref{ee}.
We first consider a normal beam splitter (shaded block in Fig. \ref{Stub}) in which lead 3
is kept at a potential $eV$ and leads 1 and 2 are grounded so that the same
bias voltage is established between 3 and 1, and 3 and 2.
The two conductors, which connect the beam splitter to the leads 1 and 2,
are assumed to be normal-metallic and perfectly transmissive, so that the S-matrix of the
system for $\theta_1=\theta_2=0$ is equal to the S-matrix of the beam splitter,
which reads~\cite{buttiker84}
\begin{equation}
S=\left( \begin{array}{ccc}
-(a+b) & \sqrt{\epsilon} & \sqrt{\epsilon} \\ \sqrt{\epsilon} & a & b\\ \sqrt{\epsilon} & b & a
\end{array} \right) .
\label{S-stub}
\end{equation}
In this parametrization of a symmetric beam splitter $a=\pm(1+\sqrt{1-2\epsilon})/2$,
$b=\mp(1-\sqrt{1-2\epsilon})/2$ and $0<\epsilon <1/2$.
For arbitrary angles $\theta_1$ and $\theta_2$, the S-matrix is obtained rotating
the quantization axis in the two conductors independently by applying the
transformation $S_{\theta_1,\theta_2}={\cal U} S {\cal U}^{\dagger}$,
where ${\cal U}$
is the rotation matrix given by:
\begin{equation}
{\cal U}=\left( \begin{array}{ccc} \mathbb{I} & 0 &0\\0&U_{\theta_1}&0\\
0&0&U_{\theta_2} \end{array} \right)
\end{equation}
and $U_{\theta}$ is defined in Eq.(\ref{uteta}).
This procedure is valid as long as no back scattering is present in the conductors.
The probability distributions are given by Eqs.~(\ref{stp}) and (\ref{FCSt}) where the state
$\ket{\psi}$ is now factorisable:
\begin{equation}
\ket{\psi}=a^{\dagger}_{1\uparrow}(E) a^{\dagger}_{1\downarrow}(E)
\ket{0}
\label{initial}
\end{equation}
in the energy range $0<E<eV$.
Analogously to what was done in Section \ref{ee}, when both analyzers are present 
we set $\sigma=\sigma'=\uparrow$. When only one analyzer is present,
however, one has to use the correct characteristic
function of Eq.(\ref{chi_s1}), since one of the two additional correlators does
not vanish. Namely,
$\langle \hat{N}_{\text{O}}^{1\uparrow} \hat{N}_{\text{O}}^{1\downarrow} \rangle=
\epsilon^2$
and $\langle \hat{N}_{\text{O}}^{1\uparrow} \hat{N}_{\text{O}}^{1\downarrow} 
\hat{N}_{\text{O}}^{2\uparrow} \rangle=0$, when the upper analyzer,
for example, in Fig. \ref{Stub} is removed.
For the other expectation values we get:
\begin{equation}
\bra{\psi}\hat{N}_{\text{O}}^{1\uparrow}\ket{\psi}=
\bra{\psi}\hat{N}_{\text{O}}^{2\uparrow}\ket{\psi}=\epsilon ,
\label{n1s}
\end{equation}
\begin{equation}
\bra{\psi}\hat{N}_{\text{O}}^{1\downarrow}\ket{\psi}=\epsilon ,
\label{n2s}
\end{equation}
\begin{equation}
\bra{\psi}\hat{N}_{\text{O}}^{1\uparrow}\hat{N}_{\text{O}}^{2\uparrow} \ket{\psi}=
\epsilon^2 \sin^2 \left( \frac{\theta_1-\theta_2}{2}\right)
\label{n3s}
\end{equation}
and
\begin{equation}
\bra{\psi}\hat{N}_{\text{O}}^{1\downarrow}\hat{N}_{\text{O}}^{2\uparrow} \ket{\psi}=
\epsilon^2 \cos^2 \left( \frac{\theta_1-\theta_2}{2}\right) ,
\label{n4s}
\end{equation}
obtaining the joint probability distributions reported in Appendix \ref{PR}.
The above number operator expectation values are equal to
the case of the entangler when $\epsilon$ is replaced by $T/2$, whereas the cross-terminal
correlators are equal in the two cases if $\epsilon$ is replaced
with $T/\sqrt{2}$. From this follows that the characteristic functions for the beam splitter
possess the same dependence on the angle difference as the corresponding characteristic 
functions for the entangler (Section \ref{ee}) but have a different structure as far 
as scattering probabilities are concerned. In particular, as expected~\cite{levitov93},
the cross-correlations vanish when the two angles are equal.
On the contrary,
when the angle difference 
is $\pi$ cross-correlations are maximized.
Furthermore, when only one analyzer is present
the characteristic function shows no dependence on the angle, but it is
not factorisable, in contrast to the case of the entangler.
As a result, the single terminal probabilities, given by Eq.(\ref{stp}), are
equal in the two cases provided that $\epsilon$ is replaced
with $T/2$.
The joint probabilities for
$Q_1=Q_2=M$ are equal in the two cases if $\epsilon$ is replaced
with $T/\sqrt{2}$ (however, this replacement  is not valid in general for joint
probabilities with $Q_1,Q_2\ne M$):
\begin{equation}
P^{\theta_1,\theta_2}(M,M)=\left[ \epsilon^2 \sin^2{\left(\theta_1
-\theta_2 \over 2\right)} \right]^M
\label{qq1}
\end{equation}
\begin{equation}
P^{\theta_1,-}(M,M)=\epsilon^{2M} 
\label{qq2}
\end{equation}

The no-enhancement assumption is verified when
\begin{equation}
\epsilon \leq \frac{1}{2}
\frac{2^{Q \over M-Q}-1}{2^{Q \over M-Q}-\frac{1}{2}} ,
\end{equation}
which equals the condition of Eq.~(\ref{condition}) once $\epsilon$ is replaced
with $T/2$.
Let us first consider the case for which $Q_1=Q_2=M$.
We obtain an important result: the CH inequality is
violated for the same set of angles found for the case of the entangler,
although to a lesser extent, since the prefactors in Eqs.~(\ref{qq1}) and (\ref{qq2})
now varies in the range $0\leq \epsilon ^{2M}\leq \frac{1}{4^M}$.
In particular, in the simplest case of $M=1$, corresponding to injecting
a single pair of electrons, the maximum violation corresponds to
${\cal S}_{CH}=\frac{\sqrt{2}-1}{4}$, which is a half of the value for the entangler.
Furthermore, the plot in Fig. \ref{Srid:Q=M} is also valid in the present case
with $\overline{{\cal S}}_{CH}$ defined as $\overline{{\cal S}}_{CH}={\cal S}_{CH}/\epsilon^{2M}$,
{\em i.e.} by replacing $T/\sqrt{2}$ with $\epsilon$.
This means that a geometry like that of the
beam splitter enables to detect violation of CH inequality without any need to resort to interaction
processes to produce entanglement.

\begin{figure}
\begin{center}
\epsfig{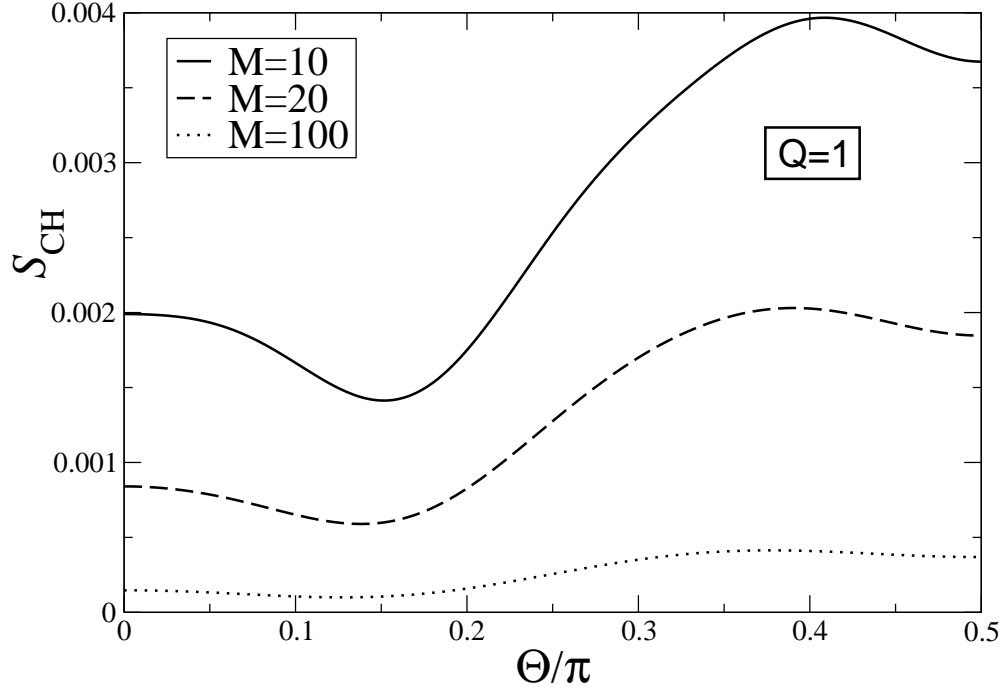}
\end{center}
\caption{The quantity ${\cal S}_{CH}$ for a normal beam splitter is plotted as a function
of the angle $\Theta$ for three values of $M=eVt/h=10,20,100$ when $Q=1$.
Interestingly, ${\cal S}_{CH}$ is positive for every angle and its maximum value
decreases like $1/M$.
}
\label{S:angle-stub}
\end{figure}

Also here we consider the case for which
$Q_1=Q_2\equiv Q<M$, where interesting differences
with respect to the case of the entangler are found.
i) We find that the violation of the CH inequality is in general weaker, meaning that
the absolute maximum value of ${\cal S}_{CH}$ is smaller than in the ideal case of
the entangler.
ii) The weakening of the violation with increasing $M$ is determined by
the suppression of the probability by the prefactor $(\epsilon^2)^{Q1+Q2}$.
Remarkably, the maximum value of ${\cal S}_{\text{max}}$ decreases like
$1/M$, therefore even slower than for the ideal case.
iii) Violations occur only for values of $Q$ close to 1, even for large values of $M$:
to search for violations one has to look at single- or few-pair
probabilities and therefore, because of the no-enhancement assumption, to
small transmissions $\epsilon$.
iv) Interestingly, for $Q=1$ the quantity ${\cal S}_{CH}$ is positive for any angles,
although the largest values correspond to $\Theta$ close to $\pi/2$
(see Fig. \ref{S:angle-stub}).
We do not find any
relevant variation, with respect to the discussion in paragraph \ref{ee},
for probabilities relative to $Q_1\ne Q_2$.

It is easy to convince oneself that the final state calculated
from the initial one (\ref{initial}) using the S-matrix (\ref{S-stub})
contains an entangled part. In Ref. \cite{bose02} this fact was already
noticed, but for an incident state composed by a single pair of particles
impinging from the two entering arms of a beam splitter. For mesoscopic 
conductors, entanglement without interaction for electrons injected from
a Fermi sea
has been discussed by Beenakker 
{\em et al}~\cite{beenakker03}.

\subsection{Superconducting beam splitter}
\label{ss}

\begin{figure}
\begin{center}
\epsfig{figure=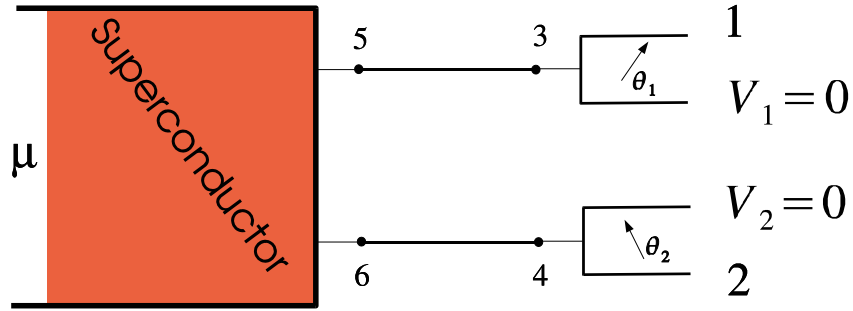,width=0.7 \textwidth}
\end{center}
\caption{Setup of a realistic system consisting of a superconducting beam splitter (shaded
region) for testing the CH inequality.
Bold lines represent two conductors of transmission probability $T$.
The superconducting condensate electrochemical potential is set to $\mu$, while
terminals 1 and 2 are grounded.
}
\label{sup_stub}
\end{figure}

In many proposals superconductivity has 
been identified as a key ingredient for the creation of entangled pairs of electrons.
The idea is to extract the two electrons which compose a Cooper pair (a pair of
spin-entangled electrons) from two spatially separated terminals.
Here we showed that it is not necessary to have superconducting correlations.
Nevertheless, in view of the recent interest in entanglement created by
pairing correlations,
it is 
useful to analyze also the case of a superconducting beam
splitter~\cite{boerlin02,samuelsson02} depicted in Fig. \ref{sup_stub}, 
which consists of a superconducting lead (with condensate chemical potential 
equal to $\mu$) in contact with two normal wires.
The wires are then connected to two leads
attached to reservoirs kept at zero potential.
This is basically what is obtained by replacing the entangler of Fig.~\ref{Ent}
by a superconducting lead with two terminals.

The system can be decomposed into two subsystems: on the left-hand-side 
of Fig.~\ref{sup_stub} we place the superconducting
slab attached two normal terminals (5 and 6)
characterized by a
reflection amplitudes matrix $R'_\text{s}$ defined, in terms of the
particle operators, by:
\begin{equation}
\hat{\phi}_{j\alpha\sigma}(E)=\sum_{k=5,6}\sum_{\beta=e,h}\sum_{\sigma'=\uparrow,\downarrow}
\left[ R'_\text{s}(E)\right]_{j\alpha\sigma,k\beta\sigma'}~
\hat{a}_{k\beta\sigma'}(E) .
\end{equation}
Here $j=5,6$ and the additional indexes $\alpha$ and $\beta$ refer to the particle-hole
degree of freedom, in particular $\alpha=e$ for particles and $\alpha=h$
for holes and $[\ldots ]_{j\alpha\sigma,k\beta\sigma'}$ represents the
specified element of the matrix.
Note that $R'_\text{s}$ is block diagonal in spin indexes so that
\begin{equation}
R'_\text{s}=\left( \begin{array}{cc}
\mathcal{R}'&0\\0&\mathcal{R}'
\end{array}\right)
\end{equation}
with
\begin{equation}
\left( \begin{array}{c}
\hat{\phi}_{5e\uparrow}\\ \hat{\phi}_{5h\downarrow}\\
\hat{\phi}_{6e\uparrow}\\ \hat{\phi}_{6h\downarrow}
\end{array}\right)=\mathcal{R}'
\left( \begin{array}{c}
\hat{a}_{5e\uparrow}\\ \hat{a}_{5h\downarrow}\\
\hat{a}_{6e\uparrow}\\ \hat{a}_{6h\downarrow}\\
\end{array}\right),\qquad \qquad 
\mathcal{R}'=\left( \begin{array}{cccc}
\rho_{ee}&\rho_{ph}&\tau_{ee}&\tau_{eh}\\
\rho_{he}&\rho_{hh}&\tau_{he}&\tau_{hh}\\
\tau'_{ee}&\tau'_{eh}&\rho'_{pp}&\rho'_{eh}\\
\tau'_{he}&\tau'_{hh}&\rho'_{hp}&\rho'_{hh}
\end{array}\right) ,
\end{equation}
where $\rho_{ee}$ ($\rho_{hh}$) is the normal reflection
amplitude for particles
(holes) in terminal 5, $\rho_{eh}$ ($\rho_{he}$) is the Andreev reflection
for a hole to evolve into a
particle (particle to evolve into a hole) in terminal 5.
$\tau_{ee}$ ($\tau_{hh}$) is the normal transmission
amplitude for particles
(holes) to be transmitted from terminal 5 to terminal 6,
$\tau_{eh}$ ($\tau_{he}$) is the Andreev transmission
amplitude for holes (particles) in terminal 5 to be transmitted
in terminal 6 as particles (holes).
Primed amplitudes refer to reflections occurring in lead 6 and transmissions
from lead 6 to lead 5.

On the right-hand-side of Fig.~\ref{sup_stub}
we have the subsystem composed of
two identical decoupled conductors characterized by the $16\times 16$
scattering matrix
\begin{equation}
S_{\text{c}}=\left( \begin{array}{cc} R_{\text{c}} & T'_{\text{c}} \\
T_{\text{c}} & R'_{\text{c}} 
\end{array}\right) .
\label{essep}
\end{equation}
The four submatrices in Eq.(\ref{essep}) are block diagonal in spin space,
for example $R_{\text{c}}$ can be written as:
\begin{equation}
R_{\text{c}}=\left( \begin{array}{cc}
R_{\text{c}}^{\uparrow}&0\\0&R_{\text{c}}^{\downarrow} ,
\end{array}\right)
\end{equation}
where $R_{\text{c}}^{\uparrow}$ is a diagonal matrix defined by
\begin{equation}
\left( \begin{array}{c}
\hat{\phi}_{3e\uparrow}\\ \hat{\phi}_{3h\downarrow}\\
\hat{\phi}_{4e\uparrow}\\ \hat{\phi}_{4h\downarrow}
\end{array}\right)=R_{\text{c}}^{\uparrow}
\left( \begin{array}{c}
\hat{a}_{3e\uparrow}\\ \hat{a}_{3h\downarrow}\\
\hat{a}_{4e\uparrow}\\ \hat{a}_{4h\downarrow}\\
\end{array}\right),\qquad \qquad 
R_{\text{c}}^{\uparrow}=\left( \begin{array}{cccc}
r_{3e\uparrow}&0&0&0\\
0&r_{3h\downarrow}&0&0\\
0&0&r_{4e\uparrow}&0\\
0&0&0&r_{4h\downarrow}
\end{array}\right) .
\end{equation}
$R_{\text{c}}^{\downarrow}$ is defined like $R_{\text{c}}^{\uparrow}$
exchanging $\uparrow$ with $\downarrow$, whereas $T_{\text{c}}^{\sigma}$
is defined similarly to $R_{\text{c}}^{\sigma}$ replacing
$r_{3\alpha\sigma}$ with $t_{1\alpha\sigma}$ and
$r_{4\alpha\sigma}$ with $t_{2\alpha\sigma}$.
The matrices $R_{\text{c}}^{\prime\sigma}$ and $T_{\text{c}}^{\prime\sigma}$ are
defined analogously using the amplitudes $r_{1\alpha\sigma}$,
$r_{2\alpha\sigma}$, $t'_{1\alpha\sigma}$ and $t'_{2\alpha\sigma}$.
The spin quantization axis of the two wires can be rotated independently
as in paragraph \ref{ee} by applying the transformation 
$S_{\theta_1,\theta_2}={\cal U} S_{\text{c}} {\cal U}^{\dagger}$, where ${\cal U}$ is defined
in Eq.~(\ref{tra}),
obtaining the scattering matrix
\begin{equation}
S_{\theta_1,\theta_2}=\left( \begin{array}{cc} \tilde{R}_{\text{c}} & \tilde{T}'_{\text{c}} \\
\tilde{T}_{\text{c}} & \tilde{R}'_{\text{c}} \end{array}\right)~.
\end{equation}
The overall matrix of reflection amplitudes is calculated by composing
the scattering matrices relative to the two subsystems \cite{datta95}:
\begin{equation}
R'_{\text{tot}}=\tilde{R}'_{\text{c}} +\tilde{T}_{\text{c}} \left[ \mathbb{I} -
R'_{\text{s}} \tilde{R}_{\text{c}}\right]^{-1} R'_{\text{s}} \tilde{T}'_{\text{c}} .
\end{equation}
where $R'_{\text{tot}}$ is defined by
\begin{equation}
\hat{\phi}_{j\alpha\sigma}(E)=\sum_{k=1,2}\sum_{\beta=e,h}\sum_{\tau=\uparrow,\downarrow}
\left[ R'_\text{tot}(E)\right] _{j\alpha\sigma,k\beta\tau}~
\hat{a}_{k\beta\tau}(E) ,
\end{equation}
with $j$ running from 1 to 2.
The characteristic function can now be calculated through Eq.
(\ref{chiEdet}) taking $R'_\text{tot}(E)$ as scattering matrix.
In the present case, where superconductivity is present, the diagonal matrix of Fermi
distribution functions is defined as
$[n_E]_{j\alpha\sigma,j\alpha\sigma}=f_{j\alpha}(E)$,
$f_{j\alpha}(E)=[1+\exp (\frac{E+\alpha\mu}{k_BT})]^{-1}$ and
$[\Lambda]_{j\alpha\sigma,j\alpha\sigma}=\exp (i\alpha\lambda_{j\sigma})$,
with $j=1,2$.
By choosing $\lambda_{1\downarrow}=\lambda_{2\downarrow}=0$ we achieve the
goal of counting excitations with spin-up component.
The case where one of the analyzers is removed,
for example in lead 1,
is implemented by setting $\lambda_{1\downarrow}=\lambda_{1\uparrow}
=\lambda_1$ and $\theta_1=0$, {\it i.e.} by counting electrons in lead 1 regardless
their spin.

In the limit of zero temperature and small bias voltage, we only need
the scattering amplitudes at the zero energy (Fermi level) so that the
overall characteristic function can be approximated like in Eq.(\ref{approx}).
We parametrize the matrix $S_{\text{c}}$ of the wires
as follows: $r_{3e\sigma}=r_{4e\sigma}=\sqrt{1-T}$,
$r_{1e\sigma}=r_{2e\sigma}=\sqrt{1-T}$,
$t_{1e\sigma}=t_{1e\sigma}=\sqrt{T}$ and
$t'_{1e\sigma}=t'_{2e\sigma}=-\sqrt{T}$, where $T$ is the wire transmission
probability of the wires.
The amplitudes relative to hole degree of freedom are determined from
the ones above by making use of the particle-hole symmetry.

Although Andreev processes are fundamental for the injection of Cooper pairs,
in the case where Andreev transmissions only are non-zero and $T=1$ the 
joint probabilities factorize in a trivial way
\begin{equation}
P^{\theta_1,\theta_2}(Q_1,Q_2)=\delta_{Q_1,2M}
\delta_{Q_2,2M} \;\;\;\;\; P^{\theta_1,-}(Q_1,Q_2)=\delta_{Q_1,2M}
\delta_{Q_2,4M}\, ,
\end{equation}
in such a way that the CH inequality is never violated.
This apparent contradiction is due to the fact that in this situation the 
scattering processes occur with unit probability, so that the condition of
locality is fulfilled.
Non-locality can be achieved by imposing $T< 1$.
In the limit $T\ll 1$ we obtain the probabilities
$P^{\theta_1,\theta_2}(Q_1,Q_2)$ and $P^{-,\theta_2}(Q_1,Q_2)$
reported, respectively, in Eqs.~(A\ref{probS}) and (A\ref{probSs}) of the Appendix,
which reduce to 
\begin{equation}
P^{\theta_1,\theta_2}(M,M)=\left[ \frac{2T^2A^6}{[A-T(A-1)]^8}\right]^M
\left[ \sin^2\left( \frac{\theta_1+\theta_2}{2}\right)\right]^M 
\label{probS1}
\end{equation}
and
\begin{equation}
P^{-,\theta_2}(M,M)=\left[ \frac{2T^2A^6}{[A-T(A-1)]^8}\right]^M
\label{probSs1}
\end{equation}
for $Q_2=Q_3=M$, with $A=1+\tau_{he}\tau^{\prime\star}_{he}$.
Eqs. (\ref{probS1}) and (\ref{probSs1}) are equal to Eqs. (\ref{prob}),
relative to the case
of an entangler, once $2T^2A^6/[A-T(A-1)]^8$ is replaced
with $T^2/2$.
From this follows that superconductivity leads to violation of the
CH inequality.
For $A=2$, {\em i.e.} perfect Andreev transmission, the quantity $2T^2A^6/[A-T(A-1)]^8$ tends to
$T^2/2$ in the limit $T\rightarrow 0$ so that the analysis of 
Section \ref{ns} relative to the case $Q_1=Q_2=M$ applies also here.

\section{Conclusions}
In mesoscopic multiterminal conductors it is possible to observe 
violations of locality in the whole distribution of the transmitted 
electrons. In this paper we have derived and discussed the CH inequality 
for the full counting electron statistics. 
In an idealized situation in which one supposes the existence of an 
{\em entangler}, we have found that the CH inequality is violated
for joint probabilities relative to an equal number of electrons 
that have passed in different terminals. This is related to the intuition that 
any violation is lost in absence of coincidence measurements.
The extent of the violation is suppressed for increasing $M$ (average number of 
injected pairs), however such a suppression does not scale exponentially with
$M$ like the probability, but instead decreases like $1/M^2$.
This means that the detection of violation does not become exponentially
difficult with increasing $M$.
For fixed transport properties we analyzed the conditions, in terms of
$M$ and number of counted electrons, for maximizing the violation of the
CH inequality.

The violation of the CH inequality could be achieved in an experiment.
Indeed we tested the CH inequality for two different realistic systems, namely
a normal beam splitter and a superconducting beam splitter.
Interestingly we find a violation even for the normal system, even though
weaker with respect to the idealized case of the entangler.
In this case the violation is again suppressed for increasing observation time,
but scales like $1/M$.
We analyzed the superconducting case in the limit of small transmissivity
and we also find a violation of the CH inequality to the same extent
with respect to the case of the entangler.

It is important to notice that the analyzers should not affect the
scattering properties of the system as in the case of ferromagnetic electrodes.
In the latter case, in fact, the probability
density of the local hidden variables would also
depend on the angles $\theta_1$ and $\theta_2$.

We believe that the results derived in this work may be of interest for the 
understanding of the statistics of electrons in mesoscopic conductors. It is 
however important to look for experimental tests of our claims. In this 
respect two possible schemes for measuring the counting statistics have been 
recently proposed in Ref. \cite{nazarov01_2}.
Since solid state devices are considered promising 
implementations for quantum computational protocols, this line of research does 
not seem interesting only from a fundamental point of view, but may be of clear relevance 
for the actual realization of solid state computers.

\begin{acknowledgments}
The authors would like to thank M. B\"uttiker, P. Samuelsson and E. Sukhorukov
for helpful discussions and C.W.J. Beenakker for comments on the manuscript.
This work has been supported by the EU (IST-FET-SQUBIT,
RTN-Spintronics, RTN-Nanoscale Dynamics).
\end{acknowledgments}

\appendix
\section{\label{PR} Probability distributions}

In this appendix we give the general expressions for the joint probability 
distributions used in the paper to discuss the CH inequality.

\subsection{Entangler}
In the case of an entangler we find
\begin{subequations}
\begin{eqnarray}
P^{\theta_1,-}(Q_1,Q_2)=\frac{T^{(Q_1+Q_2)}}{2^M} {M\choose Q_1}{M\choose Q_2}
\left( 2-T\right)^{M-Q_1}\left( 1-T\right)^{M-Q_2}\\
P^{-,\theta_2}(Q_1,Q_2)=\frac{T^{(Q_1+Q_2)}}{2^M} {M\choose Q_1}{M\choose Q_2}
\left( 1-T\right)^{M-Q_1}\left( 2-T\right)^{M-Q_2}
\end{eqnarray}
\end{subequations}
and
\begin{eqnarray}
P^{\theta_1,\theta_2}(Q_1,Q_2) &=&
 \sum_{k=\text{Max}[Q_1,Q_2]}^{\text{Min}[Q_1+Q_2,M]} 
{M\choose k} {k\choose 2k-Q_1-Q_2} {2k-Q_1-Q_2\choose k-Q_2}\times \nonumber \\
&\times &
\frac{T^{(Q_1+Q_2)}}{2^M}
\left[ 2(1-T) +T^2 \sin^2\left( \frac{\theta_1\pm\theta_2}{2}\right)\right]^{M-k}
\times \nonumber \\
& \times &
\left[ 1-T \sin^2\left( \frac{\theta_1\pm\theta_2}{2}\right)\right]^{2k-Q_1-Q_2}
\left[ \sin^2\left( \frac{\theta_1\pm\theta_2}{2}\right)\right]^{Q_1+Q_2-k}
\end{eqnarray}

\subsection{Normal beam splitter}
The joint probability $P^{\theta_1,\theta_2}(Q_1,Q_2)$ used in Section \ref{ns} is
\begin{eqnarray}
P^{\theta_1,\theta_2}(Q_1,Q_2) &=&
 \sum_{k=\text{Max}[M-Q_1,M-Q_2]}^{\text{Min}[(M-Q_1)+(M-Q_2),M]} 
{M\choose k} {k\choose M-Q_2} {M-Q_2\choose Q_1-M+k}\times \nonumber \\
&\times &
\epsilon^{(Q_1+Q_2)}
\left[ 1-2\epsilon +\epsilon^2 \sin^2\left( \frac{\theta_1-\theta_2}{2}\right)\right]
^{2M-Q_1-Q_2-k}
\times \nonumber \\
& \times &
\left[ 1- \epsilon\sin^2\left( \frac{\theta_1-\theta_2}{2}\right)\right]^{Q_1+Q_2-2M+2k}
\left[ \sin^2\left( \frac{\theta_1-\theta_2}{2}\right)\right]^{M-k}
\end{eqnarray}

\noindent
The single-analyzer joint probability $P^{-,\theta_2}(Q_1,Q_2)$ reads:
\begin{eqnarray}
P^{-,\theta_2}(Q_1,Q_2) &=& \epsilon^{(Q_1+Q_2)}
\sum_{k=0}^{Q_1}~~~~~
\sum_{l=\text{Max}[0,(Q_1-k)+(Q_2-k)]}^{\text{Min}[M-k,Q_2]} 
{M\choose k} {M-k\choose l} {k\choose k+l-Q_2} 
\times \nonumber \\
&\times &
{k+l-Q_2\choose Q_1-k} 
\left[ 1-3\epsilon +2\epsilon^2 \right]^{M-k-l}
\left[ 1- \epsilon \right]^l
\left[ 2- 3\epsilon\right]^{2k+l-Q_1-Q_2}
\end{eqnarray}
with $0\le Q_1\le 2M$ and $0\le Q_2\le M$ (note that the sum on $l$ has to be
performed only when the lower limit is less than or equal to the upper limit). 

\subsection{Superconducting beam splitter}

The joint probability $P^{\theta_1,\theta_2}(Q_1,Q_2)$ used in Section \ref{ss} is
\begin{eqnarray}\nonumber
P^{\theta_1,\theta_2}(Q_1,Q_2) &=&
\sum_{k=\text{Max}[Q_1,Q_2]}^{\text{Min}[Q_1+Q_2,M]} {M\choose k}
{k\choose 2k-Q_1-Q_2} {2k-Q_1-Q_2\choose k-Q_2}\times \nonumber \\
& \times & 
\left[ \frac{A^8}{[A-T(A-1)]^8}\right]^M
\left(\frac{2T^2}{A^2}\right)^k \times \nonumber \\
& \times & 
\left[ 
1-4T+6T^2+\frac{2T^2}{A^2} \sin^2\left( \frac{\theta_1+\theta_2}{2}\right)
\right]^{M-k}\times \nonumber \\ 
& \times & 
\left[ \sin^2\left( \frac{\theta_1+\theta_2}{2}\right)\right]^{Q_1+Q_2-k}
\left[ \cos^2\left( \frac{\theta_1+\theta_2}{2}\right)\right]^{2k-Q_1-Q_2}
\label{probS}
\end{eqnarray}
where $A=1+\tau_{hp}\tau^{\prime\star}_{hp}$.

\noindent
The single-analyzer joint probability $P^{-,\theta_2}(Q_1,Q_2)$ reads:
\begin{equation}
P^{-,\theta_2}(Q_1,Q_2)={M\choose Q_1} {Q_1\choose Q_2}
\left( \frac{A^8}{[A-T(A-1)]^8}\right)^M \left( \frac{2T^2}{A^2}\right)^{Q_1}
[1-4T+6T^2]^{M-Q_1}
\label{probSs}
\end{equation}
for $Q_1\ge Q_2$ and $P^{-,\theta_2}(Q_1,Q_2)=0$ for $Q_1< Q_2$.

\newpage

\begin{thebibliography}{99}
\bibitem{footnote}
	The idea behind these approaches is to use the superconductor as a
 	source of singlet pairs (the Cooper pairs).
\bibitem{footnote2} 
	A central issue of these works is the ability to perform coincident 
	measurements which can be achieved either by performing short-time 
	measurement~\protect\cite{kawabata00,chtchelkatchev02} or by operating in the tunneling 
	limit~\protect\cite{samuelsson03,beenakker03}.
\end{thebibliography}
\end{document}